\documentclass[aps,prb,reprint,superscriptaddress]{revtex4-1}
\usepackage{epsf}
\usepackage{graphicx}
\usepackage{gensymb}
\usepackage{natbib}
\usepackage{float}
\usepackage{appendix}
\usepackage{gensymb}
\usepackage{upgreek}

\begin{document}
\title{Magnetic breakdown and charge density wave formation: a quantum oscillation study of the rare-earth tritellurides}

\author{P. Walmsley}
\affiliation{Department of Applied Physics and Geballe Laboratory for Advanced Materials, Stanford University, Stanford, California 94305, USA}
\affiliation{Stanford Institute of Energy and Materials Science, SLAC National Accelerator Laboratory, 2575 Sand Hill Road, Menlo Park 94025, CA 94305, USA.}

\author{S. Aeschlimann}
\affiliation{Department of Applied Physics and Geballe Laboratory for Advanced Materials, Stanford University, Stanford, California 94305, USA}
\affiliation{Stanford Institute of Energy and Materials Science, SLAC National Accelerator Laboratory, 2575 Sand Hill Road, Menlo Park 94025, CA 94305, USA.}
\affiliation{Institute of Physical Chemistry, Johannes Gutenberg-University Mainz, Duesbergweg 10-14, 55099 Mainz, Germany}
\affiliation{Graduate School Materials Science in Mainz, Staudingerweg 9, 55128, Mainz, Germany}

\author{J. A. W. Straquadine}
\affiliation{Department of Applied Physics and Geballe Laboratory for Advanced Materials, Stanford University, Stanford, California 94305, USA}
\affiliation{Stanford Institute of Energy and Materials Science, SLAC National Accelerator Laboratory, 2575 Sand Hill Road, Menlo Park 94025, CA 94305, USA.}

\author{P. Giraldo-Gallo}
\affiliation{Department of Physics, Universidad de Los Andes, Bogot\'{a}, Colombia}

\author{S. C. Riggs}
\affiliation{National High Magnetic Field Laboratory, Tallahassee, Florida 32310, USA}

\author{M. K. Chan}
\affiliation{Los Alamos National Laboratory, Los Alamos, New Mexico 87545, USA}

\author{R. D. McDonald}
\affiliation{Los Alamos National Laboratory, Los Alamos, New Mexico 87545, USA}

\author{I. R. Fisher}
\affiliation{Department of Applied Physics and Geballe Laboratory for Advanced Materials, Stanford University, Stanford, California 94305, USA}
\affiliation{Stanford Institute of Energy and Materials Science, SLAC National Accelerator Laboratory, 2575 Sand Hill Road, Menlo Park 94025, CA 94305, USA.}

\date{\today}

\begin{abstract}

The rare-earth tritellurides ($R$Te$_3$, where $R$ = La, Ce, Pr, Nd, Sm, Gd, Tb, Dy, Ho, Er, Tm, Y) form a charge density wave state consisting of a single unidirectional charge density wave for lighter $R$, with a second unidirectional charge density wave, perpendicular and in addition to the first, also present at low temperatures for heavier $R$. We present a quantum oscillation study in magnetic fields up to 65\,T that compares the single charge density wave state with the double charge density wave state both above and below the magnetic breakdown field of the second charge density wave. In the double charge density wave state it is observed that there remain several small, light pockets with the largest occupying around 0.5\% of the Brillouin zone. By applying magnetic fields above the independently determined magnetic breakown field, the quantum oscillation frequencies of the single charge density wave state are recovered, as expected in a magnetic breakdown scenario. Measurements of the electronic effective mass do not show any divergence or significant increase on the pockets of Fermi surface observed here as the putative quantum phase transition between the single and double charge density wave states is approached. 

\end{abstract}

\maketitle


\section{Introduction}

\begin{figure}
\includegraphics[width=\columnwidth]{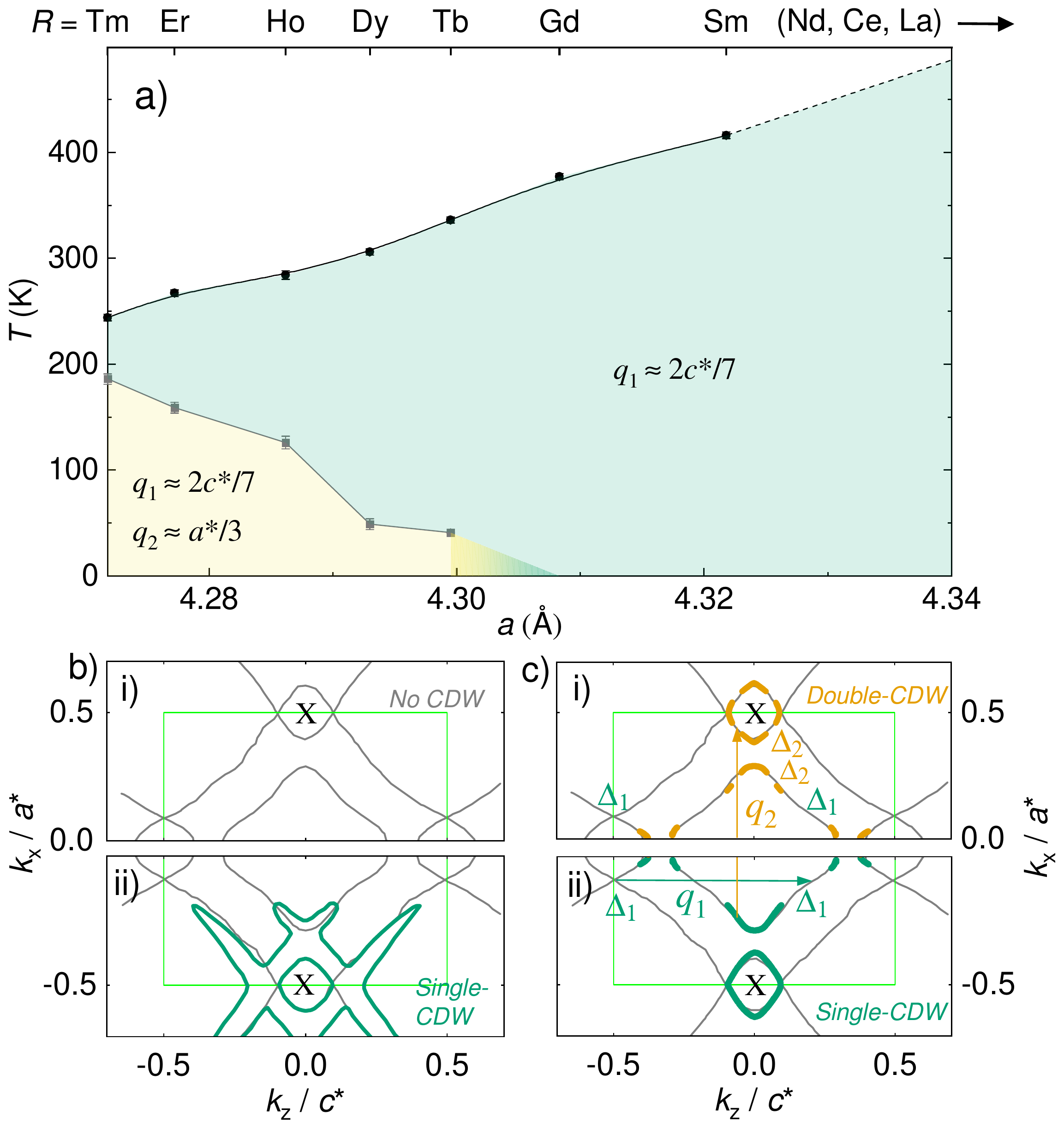}
\caption{a) Phase diagram of $R$Te$_3$ shown without the low-temperature magnetic phases (see App. \ref{APP:Magnetism}) \cite{Ru2008b, Banerjee2013} as a function of in-plane lattice parameter $a$. The top axis marks the specific rare-earth ($R$) ions that yield these lattice parameter values. There are two unidirectional, incommensurate CDWs with $q_1 \approx$ 2/7$c^*$ and $q_2\approx$ 1/3$a^*$. Both are present at low temperatures in the heaviest $R$ (shortest $a$, Tm - Tb), but just $q_1$ for lighter $R$ (longer $a$, Gd, Sm, Nd, Ce, La). The compounds where $R$=La, Ce and Nd are known to have a unidirectional CDW ordering with $q_1$ that onsets at temperatures above those measured. b) \& c) A 2-D tight-binding model that captures the essential structure of the Fermi surface of $R$Te$_3$ is shown in gray, with the first Brillouin zone shown as a light green box. Note that this model ignores a subtle $b$-axis warping and bilayer splitting. The full Fermi surface would be reflected across the $k_x=0$ line. In b.ii) a portion of the Fermi surface observed by ARPES measurements \cite{Brouet2008} in the single-CDW state is sketched in green (shown in the unfolded zone). c) Illustration of the remaining portions of the unfolded Fermi surface resolved by ARPES once gaps $\Delta_1$ and $\Delta_2$ have opened\cite{Brouet2004, Brouet2008, Moore2010}. c.i) shows this for the double-CDW state (orange) and c.ii) for the single-CDW state (green) with the CDW vectors $q_1$ and $q_2$ illustrated by the green and orange arrows respectively. Figures b) \& c) are adapted from \citet{Brouet2008} and \citet{Moore2010}.}
\label{FIG:PDFS}
\end{figure}


The Fermiology of compounds that harbour charge-density wave (CDW) order has attracted renewed interest due to the discovery of CDW order in several cuprate high-temperature superconductors\cite{Wu2011, Laliberte2011, Ghiringhelli2012, Chang2012, Achkar2012, Comin2014, daSilvaNeto2014, Tabis2014}. The results of quantum oscillation studies in the cuprates appear to be consistent with a divergence of the electronic effective mass, $m^*$, on approach to optimal doping and possibly also on the very underdoped region approaching the Mott transition \cite{Ramshaw2015, Chan2016, Barisic2013}. In both cases this effect is coincident with a dome of CDW order\cite{Keimer2015}. Close to optimal doping it remains unclear as to whether this divergence occurs as a result of a CDW quantum critical point, a quantum critical point associated with the pseudogap, or indeed some other mechanism, whereas on the underdoped side of the cuprate phase diagram, there are quantum phase transitions between CDW, spin density wave, and a Mott insulating phase that could lead to a diverging $m^*$. In addition, as a point of principle, the effect of disorder (implicit due to chemical substitution) on a unidirectional incommensurate CDW in tetragonal materials leaves only a nematic phase transition, raising the possibility that the putative CDW quantum critical point mentioned above would have a nematic character\cite{Nie2014}. As there is no clear precedent for an enhancement of $m^*$ around a CDW quantum critical point, it is clear that there is a need for a model system in which to study such a scenario. In this work we examine a promising candidate material.

The rare-earth tritellurides ($R$Te$_3$ where $R$ can be La, Ce, Pr, Nd, Sm, Gd, Tb, Dy, Ho, Er, Tm or Y) form a family of materials in which the CDW transition temperature can be smoothly tuned by lanthanide contraction without doping the system or introducing disorder\cite{Ru2008b}. As shown in the phase diagram in Fig. \ref{FIG:PDFS}, the transition temperature of the primary, unidirectional CDW order is tuned from greater than 450\,K in LaTe$_3$ down to 244\,K in TmTe$_3$, but of particular interest here is the emergence of a second, perpendicular, unidirectional CDW that first appears in a stoichiometric compound in TbTe$_3$ at 41\,K and strengthens with lanthanide contraction up to 186\,K in TmTe$_3$ \cite{Ru2008b, Banerjee2013}. As the local rare-earth moments don't seem to affect the formation of the CDW states, the choice of $R$ acts principally as chemical pressure, and as such the phase diagram can be reframed in terms of the lattice parameter as in Fig. \ref{FIG:PDFS}. This framing implies that continuous compression (expansion) of the lattice from GdTe$_3$ (TbTe$_3$) could drive the system through a quantum phase transition, potentially yielding a quantum critical point across which to search for an enhanced $m^*$.

\begin{figure*}
\includegraphics[width=\textwidth]{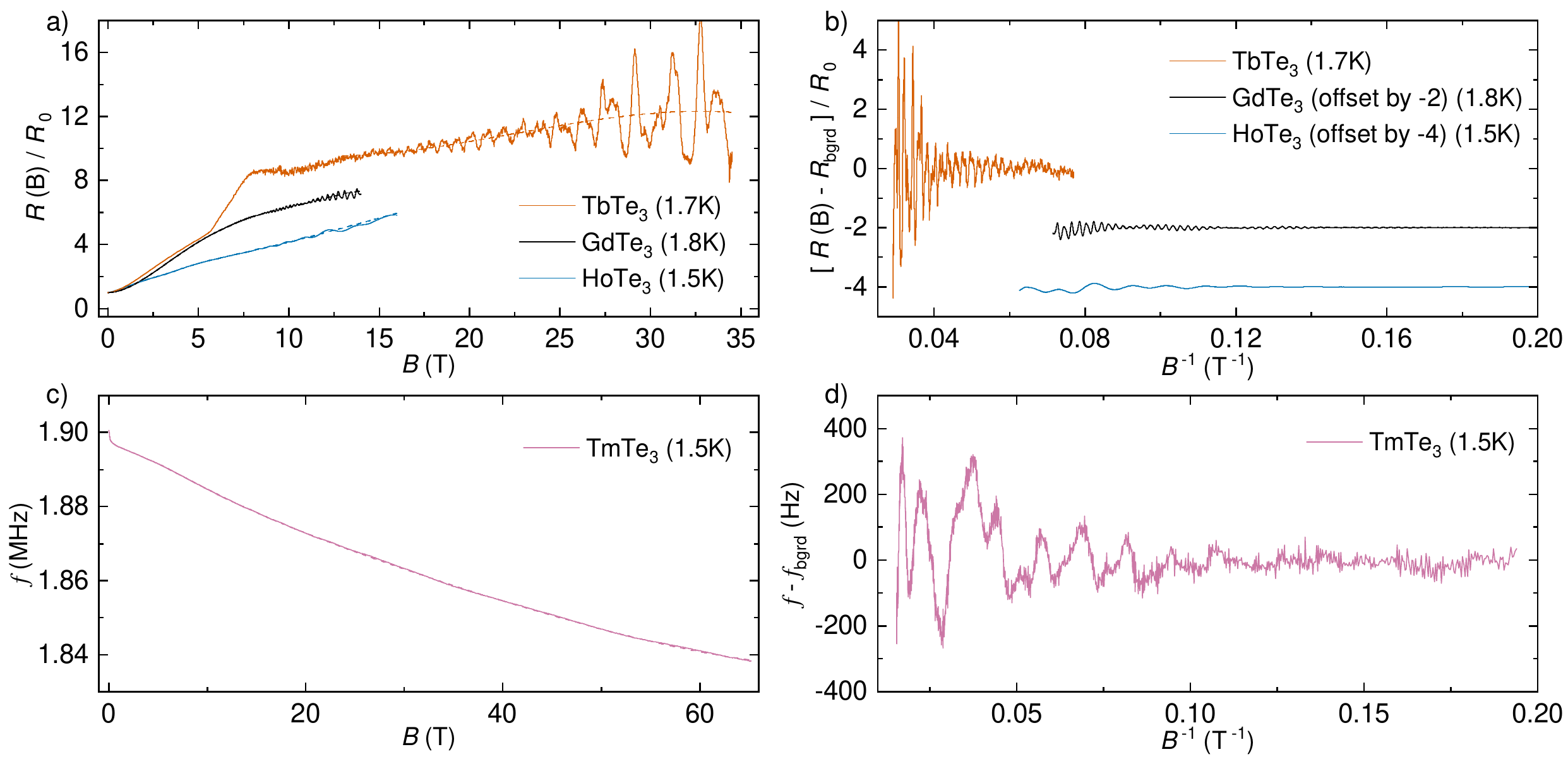}
\caption{a) $b$-axis resistance as a function of magnetic field, $R(B)$, normalised by the zero-field value $R_0$ for three representative measurements (TbTe$_3$, GdTe$_3$ and HoTe$_3$ as red, black and dark blue lines respectively), one from each DC magnet (see section \ref{SEC:Methods}). Smooth, non-oscillating background estimates, $R_{bgrd} / R_0$ for each measurement are shown as dashed lines. b) The oscillating component of the data extracted from a) by subtraction of the smooth background from the data plotted as a function of inverse field to reveal periodic oscillations indicative of quantum oscillations. c) Representative mutual inductance data shown as the mixed-down frequency of the tank circuit for TmTe$_3$, with the smooth background estimate shown as a dashed line. d) The oscillating component of the signal shown by subtraction of the non-oscillating background, $f_{bgrd}$ as a function of inverse magnetic field, again revealing periodic oscillations that are consistent with quantum oscillations.}
\label{FIG:rawdata}
\end{figure*}

The beauty of this system is that, as a stoichiometric series, disorder does not need to be introduced in order to tune the CDW phases and thus quantum oscillations can be observed down to low magnetic fields for all $R$. The Fermi surface of $R$Te$_3$ has been studied previously by ARPES in both the single and double-CDW states in CeTe$_3$ and ErTe$_3$ respectively\cite{Brouet2004, Brouet2008, Moore2010}, by positron annihilation in GdTe$_3$\cite{Laverock2005}, and by quantum oscillations in LaTe$_3$\cite{Ru2008} and GdTe$_3$\cite{Lei2020}. A 2D tight-binding model has been found to provide a good description of the Fermi surface and is shown in gray in Figs. \ref{FIG:PDFS}b) \& c) (bilayer splitting has been omitted). The Fermi surface derives from the $p_x$ and $p_z$ orbitals of the nearly-square Te net bilayer (remembering that the $b$ axis is out-of-plane in $R$Te$_3$), forming almost perpendicular, quasi 1-D sheets with weak hybridisation at their crossing points and bilayer splitting\cite{Brouet2008}. The primary CDW, with $q_1\approx 2/7c$, leaves the material metallic due to an imperfect nesting condition. ARPES and quantum oscillation studies have shown that the diamond-shaped pocket at the $\mathrm{X}$ point is unaffected by the folding, that there is likely an elongated pocket along an imperfectly nested sheet, and then another small pocket elsewhere in the zone. Figure \ref{FIG:PDFS}b.ii) illustrates a portion of Fermi surface resolved by ARPES in the single-CDW state that confirmed the survival of the $\mathrm{X}$ pocket, as well as a larger irregular pocket as a product of the zone folding. Fig. \ref{FIG:PDFS}c.ii) shows where the primary CDW gap $\Delta_1$ opens on the unfolded Fermi surface\cite{Brouet2008}.  Further ARPES data shows that as the second CDW, $q_2\approx 1/3c$, folds the Brillouin zone again, a secondary gap $\Delta_2$ also opens on the largest remaining pockets as illustrated in orange Fig.\ref{FIG:PDFS}c.i)\cite{Moore2010}.

In this study we use quantum oscillation measurements to study the folding and gapping of the Fermi surface across the implied single to double-CDW quantum phase transition. Key to understanding the data are magnetic breakdown phenomena. We observe that above the independently calculated breakdown field for the second CDW gap the quantum oscillation frequencies match those of the singly folded zone, whereas below the breakdown field only some very low frequencies remain, consistent with Fermi surface composed of just very small pockets. The quantum oscillation spectrum changes very little upon Lanthanide contraction in the single-CDW state, indicating that the choice of $R$ has a negligible effect on the area of the Fermi surface despite the size of the gap changing significantly. The temperature dependence of the quantum oscillation amplitudes show that there is no observed enhancement of $m^*$ on approach to the putative quantum phase transition.

\section{Methods}
\label{SEC:Methods}

\begin{figure*}
\includegraphics[width=17cm]{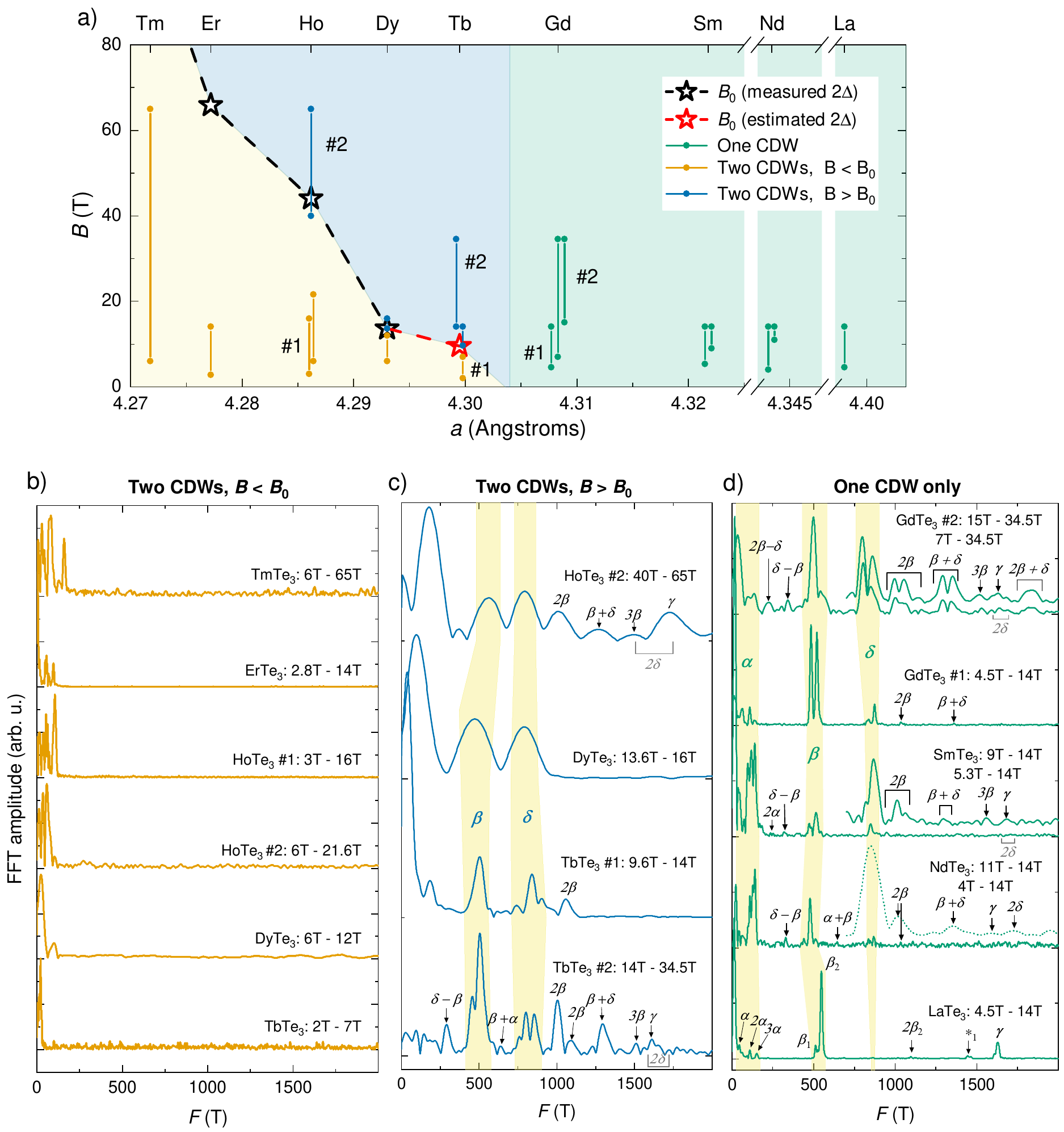}
\caption{(a) A guide showing the field ranges across which the FFTs shown in (b), (c) and (d) were obtained for each member of the $R$Te$_3$ series studied here. Each $R$ (top axis) is placed according to its $a$ lattice parameter at 300\,K (bottom axis)\cite{Malliakas2006}, noting that overlapping ranges have been symmetrically offset in $x$ for clarity. For compounds with two CDWs, the calculated characteristic breakdown field for the lower temperature (smaller gap) CDW is shown as a star. The FFTs are thus grouped into three categories; measurements in the double-CDW state with FFTs obtained from a magnetic field range below $B_0$ (orange panel and lines, panel (b)\,), measurements in the double-CDW state with FFTs derived from a magnetic field range above $B_0$ (blue panel and lines, panel (c)\,), and measurements in the single-CDW state (green panel and lines, panel (d)\,),. Note that panels (b), (c) and (d) are all plotted to the same $x$ scale for comparison. Peaks in the FFTs shown in panel (c) and (d) are labelled to identify primary frequencies from mixing frequencies and harmonics as discussed further in the main text. The primary $\alpha$, $\beta$ and $\delta$ frequencies are highlighted by yellow ribbons to more clearly show the consistency in their values as $R$ changes. In panel (d), NdTe$_3$, SmTe$_3$ and GdTe$_3$ \#2 have an additional FFT of the same data restricted to a higher range of fields to highlight high frequency components (frequencies below ~700\,T are omitted from these curves for clarity). Data in these plots were taken at fixed temperatures between 1.5\,K and 2\,K.}
\label{FIG:BreakdownMasterplot}
\end{figure*}

Single crystals of $R$Te$_3$ were grown via a self-flux technique described elsewhere\cite{Ru2006}. Quantum oscillation measurements up to 14\,T and 16\,T were performed in commercially available magnets from Quantum Design and Cryogenic Ltd respectively. Measurements to 35\,T DC fields were performed at the National High Magnetic Field Laboratory in Tallahassee and measurements to 65\,T at the pulsed-field facility at Los Alamos National Laboratory. In DC fields, quantum oscillations were measured in the electrical resistivity (Shubnikov-de Haas Oscillations) along the crystallographic $b$-axis measured by a quasi-montgomery technique. The resistivity was measured via a Stanford Research SR830 lock-in amplifier and a Princeton Applied Research Model 1900 Low Noise Transformer that added gains of 100 or 1000 owing to the very low resistance of the samples. The excitation was typically 1\,mA at 10 - 200\,Hz.  Measurements in pulsed magnetic fields utilised a mutual inductance technique \cite{VanDegrift1975}, whereby the sample is mounted on top of a flat-wound inductive coil that forms a tank circuit in combination with the coaxial line capacitance. In this configuration, changes in the resonant frequency $f$ of the tank circuit reflect changes in the average in-plane conductivity of the sample. The magnetic field was oriented parallel to the crystallographic $b$ axis (out-of-plane) in all measurements The presence of rare-earth magnetism in most members of the $R$Te$_3$ series meant that the samples had to be encased in epoxy (Devcon 5-minute epoxy) to prevent delamination and to secure the crystals against the large magnetic torques that can arise due to the large crystal field anisotropy. Encasing the samples in epoxy does not appear to significantly change their resistivity or CDW transition temperatures (constant to within 0.5\,K, around 0.1\% ), indicating that any pressure applied by the epoxy must be small.

\section{Results}

\begin{figure}
\includegraphics[width=\columnwidth]{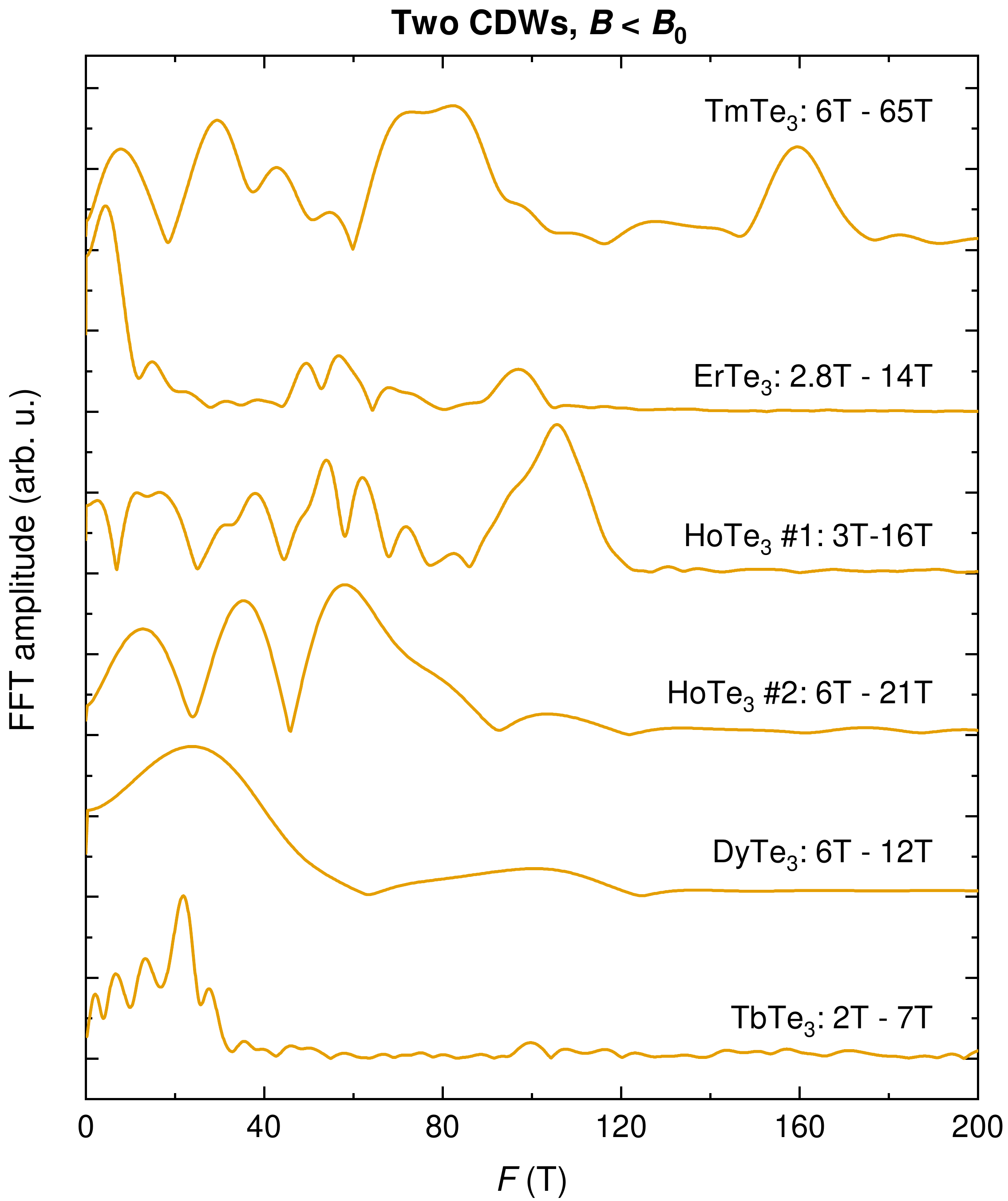}
\caption{An expanded view of the quantum oscillation frequency spectrum for $B<B_0$ in compounds with two CDWs.}
\label{FIG:2CDW_Expanded}
\end{figure}

\subsection{Magnetic Breakdown}


Representative magnetoresistance data is shown in Fig. \ref{FIG:rawdata}(a) for each of the DC magnets used in this study, with representative mutual inductance data taken in pulsed field also shown in Fig. \ref{FIG:rawdata}(c). In order to isolate the periodic oscillations associated with quantum oscillations, a smooth background (dashed lines in Figs. \ref{FIG:rawdata}(a)\&(c)\,) is removed and the data plotted versus inverse field. The low temperature magnetic phases do not appear to significantly affect the data, as shown in Appendix \ref{APP:Magnetism}, and are therefore not considered in the following analysis and discussion.

It can be seen even in the raw data in Fig. \ref{FIG:rawdata} that the dominant quantum oscillation frequencies in HoTe$_3$ and TmTe$_3$ are much lower than those in GdTe$_3$, as expected from the additional folding of the Fermi surface by the second CDW for heavier $R$ (Fig. \ref{FIG:PDFS}). TbTe$_3$ on the other hand, with two CDWs, seems to yield quantum oscillation frequencies that are more similar to GdTe$_3$, despite having a second CDW. This can be understood by considering the approximate field-scale associated with magnetic breakdown of the second CDW gap. By invoking the Blount criterion for magnetic breakdown\cite{Shoenberg1984}, $\hbar \omega_c >  E_g^2 / E_F$, where $\hbar \omega_c$ is the cyclotron frequency, $E_g$ the gap energy, and $E_F$ the Fermi energy, the magnetic breakdown field of the second CDW gap, $B_0$, can be estimated independently of the present data\footnote{Note that the magnetic breakdown reduction factor has the form $R_{MB} = (i \sqrt{P})^{l_{\nu}}(\sqrt{1-P})^{l_{\eta}}$ where $l_{\nu}$ and $l_{\eta}$ count the number of magnetic breakdown tunnelling and Bragg reflections that occur around the orbit with transmitted amplitudes $i \sqrt{P}$ and $\sqrt{1-P}$ respectively. $P = exp(-B_0/B\mathrm{cos}\theta)$ describes the magnetic breakdown probability with $B_0$ the characteristic breakdown field discussed in the main text. Without further knowledge of the Fermi surface, $\nu$ and $\eta$ cannot be known and are not discussed further here. However as we are estimating $B_0$ independently (as oppose to extracting it from fits of $R_{MB}$ to the quantum oscillation data) it remains a meaningful parameter as an estimate of the field at which appreciable magnetic breakdown should occur. Note also that as this system is quasi-2D with the magnetic field out of plane in all measurements, $\mathrm{cos}\theta = 1$ for all of the present data\cite{Shoenberg1984, Sebastian2014}.}. $E_g$ is obtained from the single particle excitation gap of the second CDW as measured by optical spectroscopy by Hu \emph{et al}\cite{Hu2014}\footnote{Gap magnitudes obtained by Hu \emph{et al.}\cite{Hu2014} are consistent with those obtained elsewhere\cite{Brouet2008, Moore2010, Pfuner2009, Pfuner2010}, but this reference is favoured here because it provides a consistent measure of the gap magnitudes for all $R$ and for both CDWs.} . Note that there has been no direct measurement of the single particle excitation gap for the second CDW of TbTe$_3$\cite{Hu2014} and so the gap magnitude has been assumed to scale proportionately with the transition temperature relative to its neighbour, DyTe$_3$. The Fermi momentum and effective mass are obtained from previous quantum oscillation measurements by Ru \emph{et al}\cite{Ru2008} to calculate $E_F$, and we use the $\beta$ frequency for this analysis because it is easily resolvable and reliably ascribed to the $\mathrm{X}$ pocket, which is known to gapped by the second CDW from ARPES measurements (as illustrated in Fig. \ref{FIG:PDFS}c).i))\cite{Ru2008, Moore2010}.

Figure \ref{FIG:BreakdownMasterplot}(a) shows the resulting values of $B_0$ as stars. While $B_0$ is not a sharp transition line, but rather the point at which the probability of tunneling via magnetic breakdown has reached $1/e$, the data can nonetheless be placed into three groups; data at fields below $B_0$ in the double-CDW state (orange in Fig. \ref{FIG:BreakdownMasterplot}), data at fields above $B_0$ in the double-CDW state (blue), and data taken in the single-CDW state (green). The quantum oscillation frequency spectrum is then  determined in each regime by taking the Fast Fourier Transform (FFT) of the oscillating component of the data as a function of inverse field confined to the relevant magnetic field range. Figure \ref{FIG:BreakdownMasterplot}(a) shows the ranges used to calculate the FFTs shown in Figures \ref{FIG:BreakdownMasterplot}(b),(c)\&(d).

Before discussing the frequency spectra in greater detail, it's worth highlighting the key trends. Compared to the data taken in the single-CDW state, only a few low frequency quantum oscillations are observed ($F \leq$ 100\,T) in the double-CDW state when the applied field is below $B_0$, which is consistent with an additional folding of the Brillouin zone and gapping of the Fermi surface. Upon applying a field greater than $B_0$ the frequency spectrum in the double-CDW state appears very similar to that in the single-CDW state. This is consistent with the magnetic breakdown scenario whereby quantum oscillation frequencies associated with an unfolded brillouin zone are recovered when the field scale exceeds the breakdown field of the hybridisation gap\cite{Harrison1996, Shoenberg1984}. The quantitative success of the independently determined $B_0$ in delineating these field scales is quite striking.

A detailed quantum oscillation study of LaTe$_3$ has been performed previously by Ru \emph{et al.} using torque and magnetic susceptibility measurements\cite{Ru2008}. It is to be expected that different techniques might be more or less sensitive to different regions of the Fermi surface, as well as observing different rules for the mixing of frequencies and appearance of harmonics. However, the primary frequencies that correspond to the real area of the Fermi surface should be reproduceable and so a comparison between techniques is an instructive place to start. In Fig. \ref{FIG:BreakdownMasterplot}(d) the frequencies observed by Ru \emph{ et al}. in LaTe$_3$, $\alpha$, $\beta_1$, $\beta_2$, $2\beta$ and $\gamma$ are indicated along with an additional frequency observed but not labelled by Ru \emph{ et al}. at around 1.45\,kT marked in \ref{FIG:BreakdownMasterplot}(d) as $*_1$. In addition, two more peaks are observed at low frequency that coincide approximately with where the second and third harmonics of $\alpha$ should be, labelled $2\alpha$, $3\alpha$. Highlighted in yellow but not marked is a very subtle peak at around 864\,T. Based on the LaTe$_3$ data along this small peak might be ignored, and it was not observed by Ru \emph{et al.} However, this subtle feature does coincide with a frequency that becomes quite clear in the other datasets in this group. As this peak is not obviously related to any harmonics, or addition and subtraction frequencies, we shall refer to it as $\delta$.

Looking at all of the traces in Figure \ref{FIG:BreakdownMasterplot}(d) the $\beta$ frequencies can be clearly seen throughout the group and are highlighted in yellow. Owing to the slight warping of the Fermi surface and bilayer splitting there are expected to be several frequencies very close to one another (Ru \emph{et al.} observe up to six for the $\beta$ frequencies) and so for comparative purposes we don't distinguish between different $\beta$ frequencies but can assume that they all originate from the same pocket of the Fermi surface. Any systematic change in these frequencies with $R$ is within the width of the group (around 440\,T to 550\,T). There is also consistently a group of low frequencies in the 40\,T to 150\,T range that are highighted in yellow and labelled $\alpha$. In GdTe$_3$ the $\alpha$ frequencies look very similar to those in LaTe$_3$, but NdTe$_3$ and SmTe$_3$ appear to have some additional frequencies in this range. It isn't obvious whether the Fermi surface has additional small pockets in these compounds or whether there are simply some addition, subtraction or harmonic frequencies that happen to be prominent in these materials. It is worth remembering that low frequencies are particularly sensitive to effects of background subtraction and windowing effects and so the relative amplitudes and precise frequencies should be interpreted with a degree of caution.

\begin{figure*}
\includegraphics[width=\textwidth]{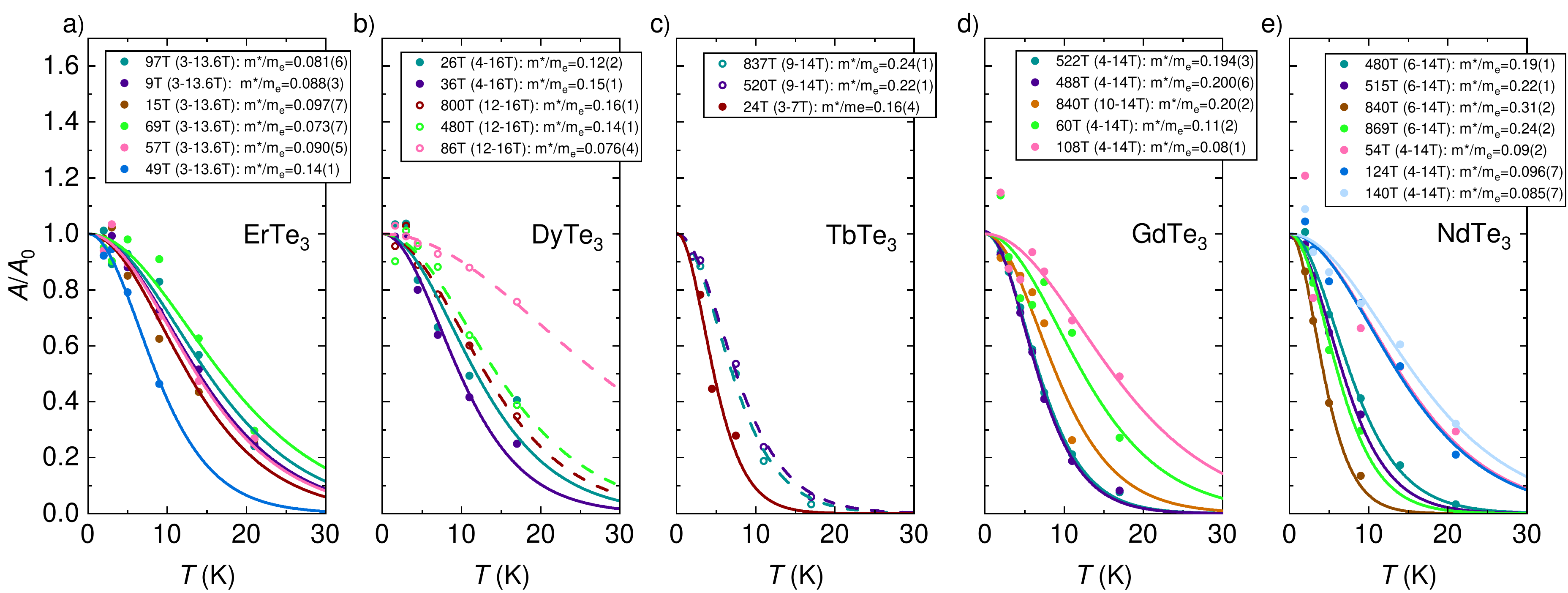}
\caption{Temperature dependence of the quantum oscillation amplitudes for primary orbits in (a) ErTe$_3$, (b) DyTe$_3$, (c) TbTe$_3$, (d) GdTe$_3$ and (e) NdTe$_3$. Fits to the Lifshitz-Kosevitch formula are shown by solid lines and yield the effective mass $m^*$. Open symbols and dashed lines correspond to breakdown frequencies. The field range used for the FFT and the frequency of the orbit are shown in the legend.}
\label{FIG:MassData}
\end{figure*}

There are two significant differences between the LaTe$_3$ spectrum and the others in this group ($R$ = Nd, Sm, Gd). The first of which is the significant reduction, or even disappearance, of the $\gamma$ frequency which is only clearly observable in LaTe$_3$. The second is the appearance in the other members of the group of another strong frequency, $\delta$, that is not clearly observable in LaTe$_3$  and doesn't appear to be a product of other fundamental frequencies. Therefore $\delta$ seems to be another fundamental frequency. It may not be a coincidence that the $\gamma$ frequency is very close to where $2\delta$ may appear, and so while the possible (weak) $\gamma$ peaks are marked on the plot, the expected location of any $2\delta$ peaks is also indicated by a grey bracket where there is ambiguity (some 3$\beta$ frequencies may also appear in this frequency range). Only LaTe$_3$ unambiguously shows a $\gamma$ peak.

The other frequencies in the spectra can be described by the addition and subtraction of the primary frequencies and their harmonics as marked on the plot. It is notable that LaTe$_3$ doesn't show addition and subtraction frequencies, with the obvious difference being that LaTe$_3$ lacks a large local moment. This is consistent with frequency mixing due to oscillations of the magnetisation. Although the other available rare-earths that could be placed in this group, Ce and  Pr, are not studied here, there is no reason to expect any variation from the broad trends described here.

The next group of FFTs, shown in Fig. \ref{FIG:BreakdownMasterplot}(b), are taken below $B_0$ in the double-CDW state. The data clearly shows that the $\beta$, $\gamma$ and $\delta$ peaks are all absent, even in data up to 65\,T in TmTe$_3$. A cluster of low frequencies remain and these are shown in more detail in Fig. \ref{FIG:2CDW_Expanded}. It is difficult to identify trends between the spectra, other than that the highest fundamental frequency is probably no higher than around 100\,T (assuming that the higher peaks unique to the TmTe$_3$ data are likely to be harmonics given that this dataset extends to considerably higher magnetic fields), which corresponds to approximately 0.5\% of the unfolded Brilluoin zone. It isn't obvious based on this data whether the $\alpha$ frequency is still present because of the density of peaks in the data being greater than our resolution. But the qualitative statement is clear; this data is consistent with a Fermi surface composed of multiple small pockets following a second folding of the Brillouin zone by the second CDW.

The final group of FFTs, representing data obtained at magnetic fields greater than $B_0$ in the double-CDW state, is shown in Fig. \ref{FIG:BreakdownMasterplot}(c). Despite having a second CDW, these datasets look much more similar to those for the single-CDW case (Fig. \ref{FIG:BreakdownMasterplot}(d)) than the double-CDW case (Fig. \ref{FIG:BreakdownMasterplot}(b)), with $\alpha$, $\beta$ and $\delta$ frequencies clearly observed. This is precisely what is expected to occur as the magnetic field scale exceeds the breakdown field of the second CDW gap. The addition and subtraction frequencies are also clearly observed in the data taken at the highest fields in TbTe$_3$ and HoTe$_3$ and marked on the plot. Taken at face value, HoTe$_3$ appears to show a $\gamma$ peak that is quite well separated from the expected position of its 2$\delta$ peak, which would suggest that this frequency is also recovered upon magnetic breakdown. However, the broad single $\delta$ peak is almost certainly representative of split peaks that are unresolved due to limited bandwidth in $1/B$. The potential $\gamma$ peak is within the range of 2$\delta$ peaks that may be anticipated given the breadth of the $\delta$ peak, and so its origin remains ambiguous. Identifying the $\alpha$ peaks clearly is problematic in these data because lower frequencies tend to be more prominent at lower fields, which by necessity are not analysed in this dataset. At intermediate fields, the frequencies of both the high and low field regimes can be expected to be present in the FFT, and so in these data we can't differentiate between the recovery of the $\alpha$ frequency by magnetic breakdown and low frequencies observed in the low field regime. Indeed, the highest field data may be beyond the quantum limit for the lowest frequencies present.

\subsection{Effective Masses}
Figure \ref{FIG:MassData} shows the temperature dependence of the FFT amplitudes for (a) ErTe$_3$, (b) DyTe$_3$, (c) TbTe$_3$, (d) GdTe$_3$ and (e) NdTe$_3$, with fits to the temperature dependent term of the Lifshitz-Kosevitch formula (that yields $m^*$) shown as lines. Dashed lines and open symbols correspond to breakdown frequencies. The data is typically more spread for the lower frequencies in part due to a greater sensitivity to the background subtraction but also due to the crowded spectrum at low frequencies. It should also be noted that many split frequencies are not individually resolvable at all temperatures, and so it's likely that many of these values are in fact an averaged value across bilayer-split Fermi surfaces or between neck and bellies of slightly warped pockets. In principle there could be some deviation from the Lifshitz-Kosevitch formula for the breakdown frequencies owing to temperature dependence of the second CDW gap, particularly in TbTe$_3$ and DyTe$_3$, but this is not resolved in the data. The error values shown in the legend are the standard errors derived from the fitting routine.

\begin{figure}
\includegraphics[width=\columnwidth]{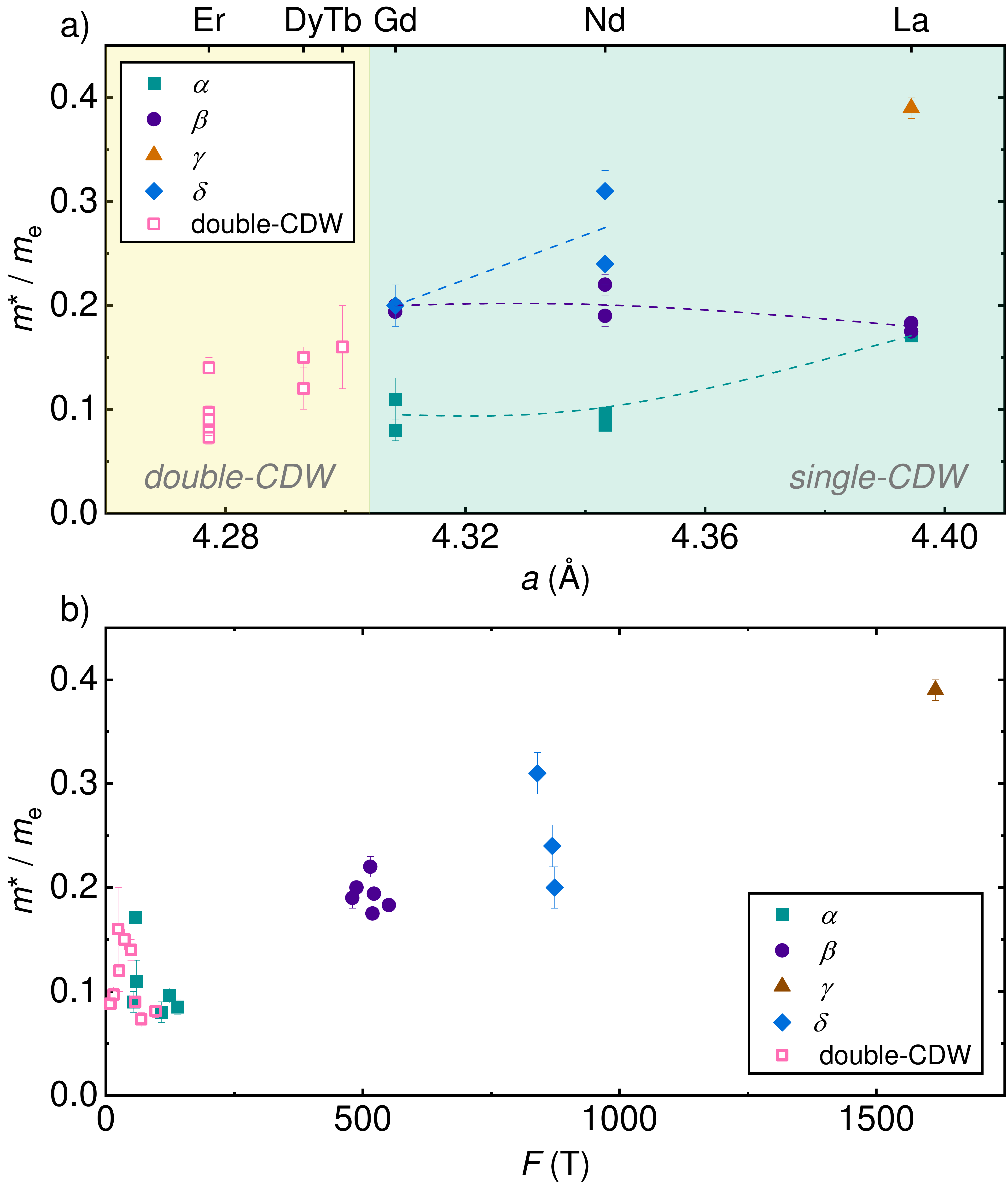}
\caption{Summary of effective mass data. (a) shows the effective masses established for various $R$, with dashed lines a guide to the eye to highlight trends upon approach to the transition to a double-CDW state. No trendline is included in the double-CDW state because it's less clear whether like orbits are being compared for different $R$. (b) shows the effective masses plotted as a function of the quantum oscillation frequency, establishing a trend that the effective mass scales with the size of the pocket.}
\label{FIG:MassSummary}
\end{figure}

The variation of $m^*$ with $R$ is shown in Fig. \ref{FIG:MassSummary}(a) with guides to the eye shown as dashed lines. There is no evidence of mass enhancement associated with the $\alpha$, $\beta$ and $\delta$ on approach to the second CDW phase, with the most clear trends actually an apparent  reduction of $m^*$ on the $\alpha$ and $\delta$ frequencies. Note that no trendline is shown in the double-CDW state because it isn't clear whether the same pocket is being tracked for each $R$ because of the crowded frequency spectrum and low resolution at low frequencies. Taking all of the data together, Fig. \ref{FIG:MassSummary}(b) shows that there is an overall trend that $m^*$ tracks with the size of the orbit such that smaller pockets of the Fermi surface are lighter than larger ones.

\section{Discussion}




The data presented here shows strong evidence that magnetic breakdown of the second CDW gap occurs with a characteristic breakdown field consistent with that calculated from independent measurements. If one dataset were viewed in isolation, it would be reasonable to question whether there were simply different Dingle terms on the higher frequency pockets to those at low frequency. This could give a similar staggered onset of frequencies with increasing magnetic field, but the systematic trends effectively eliminate this theory. Besides, it would be highly unexpected for the Fermi surface to be unchanged following the additional folding by the second CDW, as the data at fields $B>B_0$ would imply if magnetic breakdown were not invoked.

Having accounted for magnetic breakdown, it is clear from $B<B_0$ data that the Fermi surface is significantly altered by the presence of the second CDW with the largest pockets all disapearing. There remains just a series of small pockets with the largest at most 100\,T in area, or 0.5\% of the unfolded BZ. This is consistent with previous ARPES results on CeTe$_3$ and ErTe$_3$ \cite{Brouet2008, Moore2010} that show a moderately large pocket at the $\mathrm{X}$ point (thought to be the origin of the $\beta$ frequencies\cite{Ru2008}) that is ungapped by the first CDW becoming gapped by the second CDW (as illustrated in Fig.\ref{FIG:PDFS}c)). While the exact origin of the $\gamma$ and $\delta$ pockets isn't known, they also seem to be gapped by the second CDW owing to their absence from the $B<B_0$ data even up to 65\,T in TmTe$_3$. The fate of the $\alpha$ pocket is unclear as other similar frequencies appear and it isn't clear whether any of them are the original $\alpha$ frequency.

The quantum oscillation frequencies in the single-CDW state and the breakdown frequencies in the double-CDW state do not resolvably change with $R$, implying that there is no doping effect and that any changes in $q_1$ must be small so as not to significantly affect the folding of the BZ. However, the magnitude of the primary CDW gap, $\Delta_1$, does change signifcantly with $R$ and warrants discussion. Firstly, any portions of the Fermi surface that are not gapped by the primary CDW will be unaffected, which includes the $\beta$ frequencies that are thought originate from the ungapped diamond at the $\mathrm{X}$ point. The origin of the $\delta$ frequencies is not known, but the fact that they are not strongly affected by $\Delta_1$ suggests that they are either ungapped by the primary CDW, or poorly nested by $q_1$ such that their area is only weakly dependent on $\Delta_1$. This may be instructive when considering the exception, which is that the $\gamma$ frequency that is quite prominent in LaTe$_3$ is far less obvious for other $R$, while an additional primary frequency $\delta$ that isn't obviously present in LaTe$_3$ becomes strong. The $\gamma$ frequency is very close to where 2$\delta$ is expected to occur, as indicated in Fig. \ref{FIG:BreakdownMasterplot}(d), which makes it possible that $\gamma$ may only occur in LaTe$_3$. Ru \emph{et al.} argue that the $\gamma$ frequency may originate from a thin, elongated portion of the folded Fermi surface, consistent with the poor-nesting scenario put forward for the $\delta$ pockets above whereby changes in $\Delta_1$ would only tweak the tips of a long, thin pocket. Given that only one of $\gamma$ or $\delta$ seem to be prominent for a given $R$ in the single-CDW state, it would be consistent with the data to suggest that these two frequencies may originate from the same elongated piece of Fermi surface, with a change in $q_1$ or $\Delta_1$ as a function of $R$ `pinching' the long pocket in the middle and approximately halving its size as $R$ changes. Further work is required to test this hypothesis.

As discussed in the introduction, a large part of the motivation for this work was to explore $R$Te$_3$ as a model system to look for mass enhancement close to a CDW quantum critical point. In that respect this data presents a null result. The cyclotron mass associated with the $\beta$ frequency may rise moderately on approach to the putative QCP, but the $m^*$ values derived from the $\alpha$ and $\delta$ frequencies actually seem to get lighter. It is however interesting to consider why this might be the case and what the implications may be in a wider context. The first point to make is that while there is an implied quantum phase transition to the double-CDW state at a value of the lattice parameter between those of $R$=Tb and $R$=Gd, it is not known whether this phase transition would remain continuous or whether it may become first order. In the latter case, there is no reason to expect quantum critical fluctuations that would lead to mass enhancement, although some remain in the case of a weakly first order transition. Also, without having located the quantum phase transition's location in phase space exactly, it is possible that these measurements simply were not performed in close enough proximity to the quantum phase transition to be strongly influenced by it; a factor that would depend on the critical exponents. 

A second point of note is that quantum oscillations yield a cyclotron effective mass i.e. the effective mass averaged around the cyclotron orbit. As only the parts of the Fermi surface nested by $q_2$ may be expected to be renormalised then if the Fermi surface is only poorly nested the majority of the orbit may not be renormalised, thus leaving the cyclotron effective mass mostly unchanged. In support of this idea, the peaks in the Lindhard function are quite localised in $k$-space \cite{Johannes2008}, implying that the renormalised portion of the Fermi surface would form a small portion of an orbit-averaged value, however ARPES shows that a significant portion of the $\mathrm{X}$ pocket ($\beta$ frequency) is gapped, implying that the same portion of the Fermi surface is well nested. In quantum oscillation measurements, it is also always possible that there is another peak that is unobserved on which the mass enhancement is to be found, and it would be consistent with the present data if, for example, the $\gamma$ peak becomes less prominent due to an increase in its effective mass on approach to the quantum phase transition damping its amplitude. It would still remain somewhat surprising however that no mass enhancement at all would be observed on the $\beta$ pocket as it is clearly gapped and therefore must be somewhat nested by the critical boson.

An interesting trend that emerges from the data is that the effective mass scales with the frequency of the peak and hence the size of the Fermi surface pocket.  In general, small Fermi surface pockets tend to be close to band edges, and in effect folding of the BZ by a CDW introduces new band edges where the $q$ vector connects states that are slightly above or below the Fermi level. Therefore one explanation is that the states closest to the hybridisation points are lighter than those that are farther from them due to the dispersion relation in the vicinity of the hybridisation point. Alternatively, the areas of Fermi surface that are well nested by $q$ may be expected to have the largest electron-phonon coupling, and therefore also the largest m*, but these are the same states that are gapped out when the system orders, hence leaving lighter carriers at the Fermi level as the folding of the BZ succesively reduces the size of the pockets.

\section{Conclusion}
To conclude, this study presents a characterisation of quantum oscillations in the $R$Te$_3$ system that identifies magnetic breakdown of the second CDW as the origin of the observed magnetic field dependences of the quantum oscillation frequencies. The Fermi surface in the double-CDW state is observed to consist of just small pockets with a maximum area of around 0.5\% of the Brillouin zone, consistent with previous ARPES measurements. The effective mass is not observed to be renormalised close to the quantum phase transition between the single and double-CDW states for the observed orbits, and while the reason for this is not clear, several possible avenues through which to explain the absence of this effect are discussed. Although further work is required to fully understand the observed lack of mass enhancement, this work nonetheless narrows the parameters in which the search should take place in the $R$Te$_3$ family and provides a greater understanding of their Fermi surfaces.

\section{Acknowledgements}
This work was supported by the Department of Energy, Office of Basic Energy Sciences under contract DE-AC02-76SF00515. The National High-Magnetic Field Laboratory and related technical support are funded by the National Science Foundation Cooperative Agreement Number DMR-1157490 and DMR-1644779, the State of Florida and the U.S. Department of Energy. Pulse field measurements were supported by the US Department of Energy “Science of 100 tesla” BES program. S.A. is a recipient of a DFG-fellowship through the Excellence Initiative by the
Graduate School Materials Science in Mainz (GSC 266). JAWS acknowledges support as an ABB Stanford Graduate Fellow.
\section{Appendices}
\begin{appendices}

\subsection{Influence of magnetic order on quantum oscillation frequencies.}
\label{APP:Magnetism}

\begin{figure}
\includegraphics[width=\columnwidth]{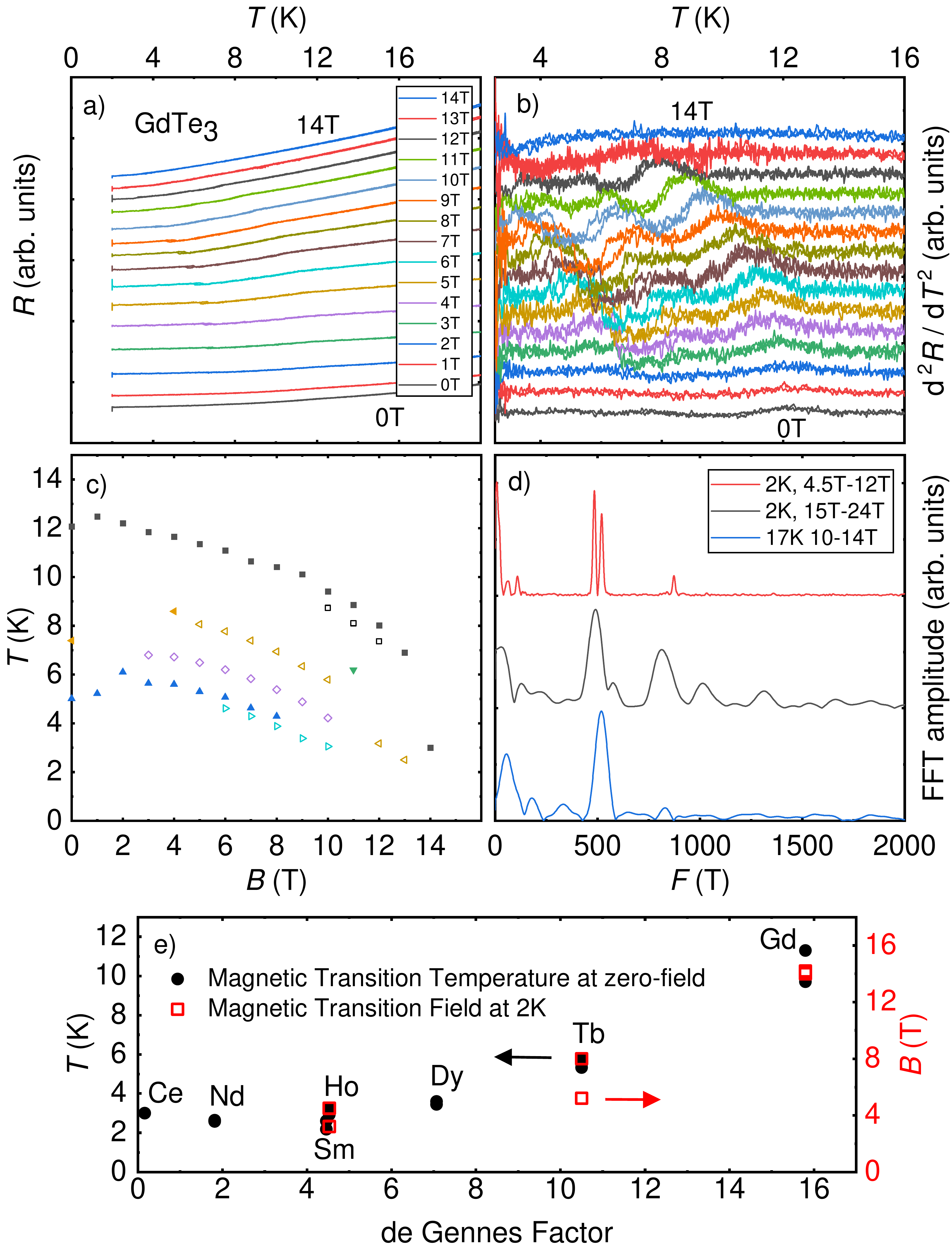}
\caption{a) $b$ axis resistivity data in fixed magnetic field ($B \parallel b$) in GdTe$_3$ taken on both warming and cooling every 1\,T between 0\,T and 14\,T (data offset for clarity). b) Second derivative in temrperature of the data in a)  (offset for clarity). c) the $B - T$ phase diagram derived from the data in b); closed symbols represent peaks in the second derivative, open symbols are taken from the centre of the thermal hysteresis loops in (b), implying a first-order phase transition. The phase boundaries seems to be approaching 0\,K just slightly above 14\,T. d) Three FFTs taken in different regions of $B-T$ phase space in GdTe$_3$; one from within the magnetic phase (2\,K, 4.5\,T - 12\,T, red line), one above the magnetic phase boundary in field (2\, K, 15\,T - 24\,T, black line), and one from above the magnetic phase in temperature (17\,K, 10-14\,T, blue line). This illustrates that any alteration of the quantum oscillation spectrum by magnetic ordering must be subtle enough not to affect our conclusions. e) the magnetic phase transitions determined previously without applied magnetic field by Ru \emph{et al.} (black circles, left axis) \cite{Ru2008c} compared to the magnetic polarisation fields observed in applied fields at 2\,K for $R$=Gd, Tb and Ho (red squaress, right axis) as a function of the de Gennes factor. Determination of the polarisation fields is discussed in the text. The polarisation fields seem to scale with the zero-field transition temperatures.}
\label{FIG:MagneticPD}
\end{figure}

The rare-earth moments in $R$Te$_3$ order at low temperatures, but do not appear to significantly influence our results. To explicitly check this, we have mapped out the magnetic phase diagram as a function of magnetic field for the relevant field orientation $B \parallel b$ for GdTe$_3$, which can be considered a `worst case' scenario as it has the highest ordering temperature and the largest de Gennes factor. Figs. \ref{FIG:MagneticPD}(a)\&(b) respectively show resistivity curves at fixed magnetic fields and their second derivatives in temperature as a means to identify features that may be associated with phase boundaries. Figure \ref{FIG:MagneticPD}(c) shows the results of this analysis, with peaks in the second derivative of the resistivity plotted as solid symbols, and thermal hysteresis loops, which may be indicative of first-order transitions, marked as open symbols. The symbols are coloured based on tracking similar looking features. The magnetic phase appears to be fully suppressed just slightly above 14\,T.

In order to test whether the magnetic phase is affecting the quantum oscillation data, FFTs (presented in Fig. \ref{FIG:MagneticPD}(d) ) were taken in the magnetic phase (2\,K, 4.5\,T - 12\,T) and outside of the magnetic phase in both higher fields (2\,K,15\,T - 24\,T) and higher temperatures (17\,K, 10\,T - 14\,T). The resultant frequency spectra are qualitatively very similar, with the variation principally due to reduced bandwidth and increased noise in the high field trace, and temperature damping of the quantum oscillation at 17\,K. There may be a small shift in the frequency of the $\delta$ peak, but not to the extent that it affects any of the discussion of our results. Panel (e) in Fig.\ref{FIG:MagneticPD} shows data from \citet{Ru2008c} showing the scaling of the magnetic phase at zero field as a function of the de Gennes factor, highlighting that Gd has the strongest magnetic ordering. It is in general difficult to determine the magnetic polarisation fields in $R$Te$_3$ from magnetoresistance data because the signal is often dominated by quantum oscillations, but estimates are shown for GdTe$_3$, TbTe$_3$ and HoTe$_3$. The polarisation field in GdTe$_3$ is estimated from Figure \ref{FIG:MagneticPD}c) with the two points being two different extrapolations to 2\,K (vertical and linear), the two points shown for TbTe$_3$ are the two kinks observable in the raw data (Figure \ref{FIG:rawdata}(a)), and in HoTe$_3$ the peaks in the first and second derivative of the contactless conductivity measurement are both shown. While this is a crude analysis, it shows that the polarisation field scales with the zero-field transition temperature thus justifying the assertion that GdTe$_3$ represents a worst-case scenario and that the majority of the data analysed here is outside of the magnetically ordered phase. We can thus expect the magnetic phase to have a limited effect for other $R$ as the data is either above the ordering field or just unaffected due to the breakdown of a comparitively small gap induced by the magnetic order.

\end{appendices}

%


\end{document}